\documentclass[fp,twocolumn]{jpsj3}
\usepackage{txfonts}
\usepackage{graphicx}
\usepackage{amsmath}
\usepackage{braket}
\usepackage{here}
\usepackage{siunitx}
\usepackage{csquotes}
\usepackage{circledsteps}
\usepackage{booktabs}
\usepackage{multirow}
\newcommand{\romani}{\text{i}}
\newcommand{\romanii}{\text{ii}}
\newcommand{\romaniii}{\text{iii}}

\title{Excited core-level dependence of entanglement between a photoelectron and \\
an emitted X-ray photon in X-ray inner-shell excitation}

\author{Ryo B. Tanaka\thanks{su23179r@st.omu.ac.jp}$^{1}$, Goro Oohata$^{2,3,4}$ and Takayuki Uozumi$^{1}$}
\inst{$^{1}$Department of Physics and Electronics, Graduate School of Engineering, \\Osaka Metropolitan University,\\ Sakai, Osaka 599-8531, Japan \\
$^{2}$Department of Physics, Graduate School of Science,\\ Osaka Metropolitan University,\\ Sakai, Osaka 599-8531, Japan \\
$^{3}$Department of Physics Nambu Yoichiro Institute of Theoretical\\ and Experimental Physics (NITEP),\\ Osaka Metropolitan University,\\ Sugimoto 3-3-138, Sumiyoshi-ku, Osaka 558-8585, Japan \\
$^{4}$Research Institute for Light-induced Acceleration System (RILACS),\\ Osaka Metropolitan University,\\ 1-2 Gakuencho, Nakaku, Sakai, Osaka, 599-8570, Japan}

\abst{We theoretically investigated how the quantum entanglement between the spin of the photoelectron and the polarization of the emitted X-ray photon depends on the excited core-level, using the 3$d\rightarrow\ $2$p$ and 3$d\rightarrow\ $3$p$ SPR-XEPECS (spin- and polarization-resolved XEPECS) processes for $\mathrm{Ti_{2}O_{3}}$-type system, and the 4$f\rightarrow\ $4$d$ SPR-XEPECS process for $\mathrm{CeF_{3}}$-type system.
In the calculation for $\mathrm{Ti_{2}O_{3}}$-type system, we used $\mathrm{TiO_{6}}$ cluster model with the full-multiplet structure of the Ti ion and the charge-transfer effect between Ti 3$d$ and ligand O 2$p$ orbitals. 
For $\mathrm{CeF_{3}}$-type system, we used ionic model with the full-multiplet structure of the Ce ion.
We found two distinct mechanisms for entanglement generation in the 3$d\rightarrow\ $2$p$ and 4$f\rightarrow\ $4$d$ cases. 
The first is generated by the spin–orbit interaction of the 2$p$ core electron, whereas the second is generated by the spin–orbit interaction of the 4$f$ valence electron and strong exchange interaction between the 4$f$ and 4$d$ electrons.
However, in the 3$d\rightarrow\ $3$p$ case with the strong 3$d-$3$p$ exchange interaction, we found that the entanglement is not generated due to the crystal field effect.
These results reveal the existence of two distinct mechanisms for entanglement generation in X-ray inner-shell excitation processes.}

\begin{document} 
\maketitle

\section{Introduction}
Quantum entanglement, which exhibits nonlocal correlations unique to quantum mechanics\cite{HORODECKI}, has been primarily studied in the low-energy optical regime for its applications in quantum information and communication technologies\cite{TOGAN,KOSAKA,TSURUMOTO,SEKIGUCHI,REISERER}.
In contrast, in the high-energy optical regime such as X-rays, recent advances in synchrotron radiation sources have stimulated active research on quantum entanglement\cite{CHOWDHURY, SHWARTZ1, SHWARTZ2, RIVERA, SOFER, STRIZHEVSKY, ZHANG, HARTLEY1, HARTLEY2}. 
Among various approaches, X-ray parametric down-conversion, which generates entangled X-ray photon pairs via a nonlinear optical effects without involving inner-shell excitation, has become the primary method in this regime\cite{SHWARTZ1, SHWARTZ2, SOFER, STRIZHEVSKY, HARTLEY1, HARTLEY2}.
In contrast, there have been very few reports on the generation of quantum entanglement via X-ray inner-shell excitation processes\cite{RTANAKA1, RTANAKA2}, and it remains unclear how the selectivity of the excited core-levels---a characteristic feature of X-ray inner-shell excitations---affects the generation and degree of the entanglement.

In this study, we treat the XEPECS process, i.e., XES (X-ray emission spectroscopy) $\&$ cXPS (core-level X-ray photoemission spectroscopy) coincidence spectroscopy\cite{STANAKA}, one of the X-ray spectroscopic processes, as a method to produce a pair of electron and photon. 
XEPECS simultaneously measures the kinetic energy of photoelectrons and the energy of X-ray photons emitted from the same atom. 
In this paper, to investigate the quantum entanglement between the spin of photoelectrons and polarization of emitted X-ray photons, we treat photoelectron spin- and emitted X-ray photon polarization-resolved XEPECS (SPR-XEPECS).

So far, we have investigated the spin and polarization entanglement in the 3$d\rightarrow\ $2$p$ SPR-XEPECS process for $\mathrm{Ti_{2}O_{3}}$ using a charge-transfer $\mathrm{TiO_{6}}$-type cluster model\cite{OKADA, UOZUMI, MATSUBARA1} considering the full-multiplet coupling effect\cite{RTANAKA2}. 
Here, in this process, a 2$p$ core electron is emitted as a photoelectron by an incident X-ray and an X-ray photon is emitted through the Ti 3$d\rightarrow\ $2$p$ radiative decay of the 2$p$ core-hole. 
As a result, we found that the generation of spin and polarization entanglement and the degree of entanglement decrease as the charge-transfer effect increases.
However, as our previous paper treated only the 2$p$ core-level excitations, how the selectivity of the excited core-levels---a distinctive feature of X-ray inner-shell excitations---affects the generation and degree of entanglement remains unclear.

To address this issue, this paper clarifies the dependence of the photoelectron spin and emitted X-ray photon polarization entanglement on the excited core-level.
In particular, we investigate the entanglement generation in the 3$d\rightarrow\ $2$p$ and 3$d\rightarrow\ $3$p$ SPR-XEPECS process for $\mathrm{Ti_{2}O_{3}}$ to examine how it changes when the excited core-level is varied from 2$p$ to 3$p$.
Furthermore, to compare with the $\mathrm{Ti_{2}O_{3}}$ system, we investigate the 4$f\rightarrow\ $4$d$ SPR-XEPECS process for $\mathrm{CeF_{3}}$, which is a $4f^{1}$ system analogous to the 3$d^{1}$ system of $\mathrm{Ti_{2}O_{3}}$.

\begin{figure}[H]
    \centering
    \includegraphics[scale=0.5]{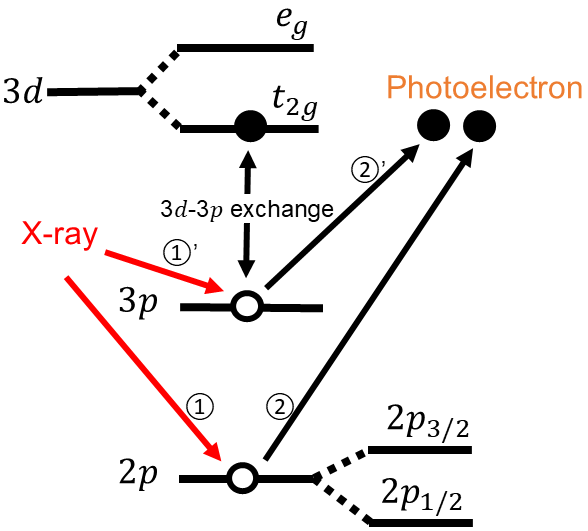}
    \caption{Schematic illustration of the $2p$ core-level excitation process ($\Circled{1}\rightarrow\Circled{2}$) and the $3p$ excitation process ($\Circled{1}^{'}\rightarrow\Circled{2}^{'}$) in $\mathrm{Ti_{2}O_{3}}$. 
    While the $2p^{5}$ state is split by $2p$ spin-orbit interaction, the $3p^{5}3d^{1}$ state is split by 3$d-$3$p$ exchange interaction.}
    \label{Fig.1}
\end{figure}

Prior to proceeding to the details of the theoretical method, the characteristic pictures of the 2$p$ and 3$p$ core-level excitations are outlined, as illustrated in Fig. \ref{Fig.1}.
In the 2$p$ core-level excitation process ($\Circled{1}\rightarrow\Circled{2}$), the 2$p^{5}$ state, formed after the emission of a photoelectron, splits into a 2$p_{1/2}$ doublet and a 2$p_{3/2}$ quartet by spin-orbit interaction.
In contrast, in the 3$p$ excitation process ($\Circled{1}^{'}\rightarrow\Circled{2}^{'}$), the exchange interaction between the 3$d$ and 3$p$ electrons dominates over the 3$p$ spin-orbit interaction, and thus, the 3$p^{5}$3$d^{1}$ state splits into a spin-triplet state with the total spin-angular momentum $S=1$ and a spin-singlet state with $S=0$ due to the exchange interaction.
In the 4$f\rightarrow\ $4$d$ process of $\mathrm{CeF_{3}}$, as in the 3$d\rightarrow\ $3$p$ process of $\mathrm{Ti_{2}O_{3}}$, the 4$d^{9}$4$f^{1}$ state splits due to the strong 4$f-$4$d$ exchange interaction.
Although there is such a similarity between 3$d\rightarrow\ $3$p$ and 4$f\rightarrow\ $4$d$ process, the crystal-field effect---one of the key effects in this paper---is significant in $\mathrm{Ti_{2}O_{3}}$, whereas it is known to be weak in $\mathrm{CeF_{3}}$\cite{NAKAZAWA1, NAKAZAWA2}.
It should be noted that the splitting into the $S=0$ and $S=1$ states occurs only in the $3d^{1}$ and $4f^{1}$ systems.
As discussed later, we find that the formation of the strong spin--polarization entangled state is deeply related to the 4$d$ core excitation at $S=0$ state in the $4f^{1}$ system.

\section{Theoretical method}
\subsection{Hamiltonian based on cluster model}
In order to investigate the dependence on the excited core-level for spin and polarization entanglement, for $\mathrm{Ti_{2}O_{3}}$, we employ a charge-transfer $\mathrm{TiO_{6}}$ cluster model with the $O_{h}$ symmetry, including the full-multiplet structure of the Ti ion.
For $\mathrm{CeF_{3}}$, on the other hand, we treat Ce as a free $\mathrm{Ce^{3+}}$ ion due to the weak crystal-field effect\cite{NAKAZAWA1, NAKAZAWA2} and the strong electronegativity of fluorine.

The Hamiltonian for $\mathrm{Ti_{2}O_{3}}$ is given by the equations,
\begin{align}
\label{ham_0}
    H_{0} =& H_{\mathrm{Ti}} + H_{\mathrm{ligand}} + H_{\mathrm{mix}} + H_{\mathrm{PE}},\\
\label{ham_ti}
    H_{\mathrm{Ti}} =& \sum_{\gamma}\epsilon_{d}(\gamma)d_{\gamma}^{\dag}d_{\gamma} + U_{dd}\sum_{\gamma>\gamma'}d_{\gamma}^{\dag}d_{\gamma}d_{\gamma'}^{\dag}d_{\gamma'}\notag\\
    &- U_{dc}\sum_{\gamma,\xi}d_{\gamma}^{\dag}d_{\gamma}(1-c_{\xi}^{\dag}c_{\xi}) + H_{\mathrm{mult}},\\ 
\label{ham_ligand}
    H_{\mathrm{ligand}} =& \epsilon_{P}\sum_{\gamma}P_{\gamma}^{\dag}P_{\gamma},\\
\label{ham_mix}
    H_{\mathrm{mix}} =& \sum_{\gamma}V(\gamma)(d_{\gamma}^{\dag}P_{\gamma} + P_{\gamma}^{\dag}d_{\gamma}),\\
\label{ham_pe}
    H_{\mathrm{PE}} =& \sum_{\vec{k},\sigma}\varepsilon_{\vec{k}} b_{\vec{k}\sigma}^{\dag}b_{\vec{k}\sigma}.
\end{align}
$H_{\mathrm{Ti}}$ in Eq. (\ref{ham_ti}) describes the Ti ion and includes 3$d$ level $\epsilon_{d}(\gamma)$, 3$d-$3$d$ Coulomb interaction $U_{dd}$ and core-hole potential $U_{dc}$ for the 3$d$ electrons in the core-excited states.
In Eq. (\ref{ham_ti}), $d_{\gamma}^{\dag}$ is the creation operator for a 3$d$ electron, and $c_{\xi}^{\dag}$ means the creation operator for a 3$p$- or 2$p$-core electron, where $\gamma$ and $\xi$ denote the combined indices of orbital and spin states. 
$H_{\mathrm{mult}}$ consists of the spin-orbit interaction for 3$d$ and core states and the Coulombic multipole parts of the 3$d-$3$d$ and the 3$d-$core interactions. 
$H_{\mathrm{ligand}}$ describes the O 2$p$ ligand molecular orbitals, which couple to the Ti 3$d$ orbitals through hybridization $H_{\mathrm{mix}}$.
$H_{\mathrm{PE}}$ is for the photoelectrons, where $\varepsilon_{\vec{k}}$ is the kinetic energy with the wave-number vector $\vec{k}$ and $b_{\vec{k}\sigma}^{\dag}$ is the creation operator for the plane wave with the $\vec{k}$ and the spin $\sigma$.
As in the previous works, the photoelectron's spin is quantized along the cluster $z$ axis\cite{RTANAKA1, RTANAKA2}.

The cluster model includes solid-state parameters such as the charge-transfer energy $\Delta (\equiv \epsilon_{d}(\gamma)-\epsilon_{P})$, the Ti 3$d$ crystal-field splitting $10Dq(=\epsilon_{d}(e_{g})-\epsilon_{d}(t_{2g}))$, the $p-d$ hybridization $V(\gamma)$, $U_{dd}$, and $U_{dc}$ as adjustable parameters. 
In this study, for the 2$p$ core-level excitation, we set $\Delta = 6.5$ eV, $10Dq = 0.5$ eV, $V(e_{g}) = 3.0$ eV, $U_{dd} = 4.5$ eV, and $U_{dc} = 5.3$ eV, following our previous work\cite{RTANAKA2}.
For simplicity, we use the same parameter values for the 3$p$ core-level excitation.

The Slater integrals $F^{k}$ and $G^{k}$, and the spin-orbit coupling constant $\zeta$ are calculated by Cowan's program\cite{COWAN}.
Here, the values of the spin-orbit coupling constants ($\zeta(2p), \zeta(3p)$) and the Slater integrals related to the exchange interaction ($G^{k}(2p,3d), G^{k}(3p,3d)$) for the representative electronic configurations $2p^{5}3d^{1}$ and $3p^{5}3d^{1}$ are listed in Table \ref{Table.1}.
As illustrated in Fig. \ref{Fig.1}, the characteristic pictures of each excitation process are quantitatively supported by the large $\zeta(2p)$ in the 2$p$ core-level excitation and the large $G^{k}(3p,3d)$ in the 3$p$ excitation, with the values listed in Table \ref{Table.1}.
Actually, the Slater integrals are reduced to 85\% for $F^{k}(3d,3d)$, $F^{2}(2p,3d)$, and $G^{k}(2p,3d)$, and to 75\% for $F^{2}(3p,3d)$ and $G^{k}(3p,3d)$, following previous studies\cite{RTANAKA2, OKADA, UOZUMI}.
In the case of $\mathrm{CeF_{3}}$, $F^{k}(4f,4f)$, $F^{k}(4d,4f)$, $G^{k}(4d,4f)$ are reduced to 80\%, 75\% and 66\%, respectively\cite{OGASAWARA}.

\begin{table}[htb]
\centering
\caption{Slater integrals and spin–orbit coupling constants (eV) for $2p^{5}3d^{1}$ and $3p^{5}3d^{1}$ configurations of Ti ion.}
\begin{tabular}{ccccccc}
\hline\hline
configuration & & \multicolumn{2}{c}{3$d-$core} & & \multicolumn{2}{c}{spin-orbit} \\
 & & $G^{1}$ & $G^{3}$ & & $\zeta(2p)$ & $\zeta(3p)$ \\
\hline
$2p^{5}3d^{1}$ & & 4.628 & 2.633 & & 3.776 &  \\
$3p^{5}3d^{1}$ & & 13.819 & 8.496 & &  & 0.434 \\
\hline\hline
\end{tabular}
\label{Table.1}
\end{table}

\subsection{Ground state of cluster system}
The ground state $\ket{g}_{\mathrm{clu}}$ of the $\mathrm{TiO_{6}}$ cluster system is described by a linear combination of three different configurations as 
\begin{align}
    \ket{g}_{\mathrm{clu}} = \alpha_{0}\ket{d^{1}}+\alpha_{1}\ket{d^{2}\underline{L}}+\alpha_{2}\ket{d^{3}\underline{L}^{2}},
\end{align}
where $\underline{L}$ means a ligand hole, and the three configurations are mixed with each other through the $p-d$ hybridization in Eq. (\ref{ham_mix})\cite{UOZUMI}.
We treat the initial state $\ket{g}$ as the direct product of the ground state $\ket{g}_{\mathrm{clu}}$ and the incident photon state $\ket{\lambda_{\mathrm{in}}}_{\mathrm{ph}}$ with linear polarization $\lambda_{\mathrm{in}}$.

Under the $O_{h}$ crystal field, the Hamiltonian in Eq. (\ref{ham_ti}) includes the crystal-field term $(21Dq)\times[C_{0}^{4}+\sqrt{5/14}(C_{4}^{4}+C_{-4}^{4})]$ acting on the 3$d$ electrons.
The terms with spherical functions $C_{\pm4}^{4}$ couple basis states whose total magnetic quantum numbers $M$ differ by 4.
As a result, four independent subspaces are formed in the $3d^{1}$ system of $\mathrm{Ti_{2}O_{3}}$, labeled by $M = -3/2, -1/2, 1/2$, and $3/2$.
In this case, the lowest-energy states in the four subspaces are degenerate under the presence of the spin-orbit interaction, exchange interaction, and crystal field, which are the key interactions in this study.
As already shown in our previous work\cite{RTANAKA2}, the spin and polarization entanglement occurs in every subspace.
In this study, however, we treat only the ground state in the subspace labeled by $M = -1/2$, in order to focus on the dependence on the excited core-level.
For $\mathrm{CeF_{3}}$, we consider the ground state of $\mathrm{Ce^{3+}}$ ion also in the $M = -1/2$ subspace.

\subsection{Entanglement evaluation using the density matrix}
To investigate the generation of spin and polarization entanglement, we calculate the density matrix\cite{RTANAKA2} 
\begin{align}
\label{eq.rho}
    \rho_{\sigma\lambda,\sigma'\lambda'}(E_{B},\omega) = \sum_{\sigma,\sigma',\lambda,\lambda'}&C_{\sigma\lambda,\sigma'\lambda'}(E_{B},\omega)\notag\\
    &\times\ket{\sigma}_{\mathrm{el}}\ket{\lambda}_{\mathrm{ph}}\bra{\sigma'}_{\mathrm{el}}\bra{\lambda'}_{\mathrm{ph}}.
\end{align}
This density matrix is obtained by a partial trace over the degrees of freedom of the cluster system from the projector constructed from the wavefunction of the SPR-XEPECS process, which is treated as a coherent second-order optical process.
Here, the density matrix consists of the coupled state of the photoelectron's spin $\ket{\sigma}_{\mathrm{el}}$ and the emitted X-ray photon's linear polarization $\ket{\lambda}_{\mathrm{ph}}$. 
In addition, the matrix element $C_{\sigma\lambda,\sigma'\lambda'}(E_{B},\omega)$ depends on the binding energy $E_{B}$ and the emitted photon energy $\omega$.  
In this paper, to evaluate the spin and polarization entanglement, we use the fidelity and the tangle based on the density matrix of Eq. (\ref{eq.rho}). 
The fidelity $F$ quantifies the similarity between the calculated $\rho_{\sigma\lambda,\sigma^{'}\lambda^{'}}$ and the so-called ideal state $\ket{\psi}$, and is defined by $F=\braket{\psi|\rho_{\sigma\lambda,\sigma^{'}\lambda^{'}}|\psi}$\cite{EDAMATSU,OOHATA}.
The tangle $T$, on the other hand, quantifies the degree of quantum entanglement\cite{WOOTTERS,COFFMAN,JAMES,MUNRO,MUNRO2}, which is given by
\begin{align}
    \label{eq.tangle}
    T = C^{2} = [\mathrm{max}(\sqrt{\Lambda_{1}}-\sqrt{\Lambda_{2}}-\sqrt{\Lambda_{3}}-\sqrt{\Lambda_{4}}, 0)]^{2}
\end{align}
using the concurrence $C$.
Here, $\Lambda_{i}\mathrm{s}\ (\textit{i}=1\sim4)$ are the eigenvalues arranged in decreasing order of the matrix
\begin{align}
    \rho_{\sigma\lambda,\sigma'\lambda'}(\sigma_{y}\otimes\sigma_{y})\rho_{\sigma\lambda,\sigma'\lambda'}^{*}(\sigma_{y}\otimes\sigma_{y}).
\end{align}
The tangle $T$ takes values from 0 to 1, and indicates higher degree of entanglement for $T$ close to 1.

\subsection{Formulation of spectral calculations}
The XPS is calculated using the spectral function for the first-order optical process as
\begin{align}
\label{eq.xps}
    F_{\mathrm{XPS}}(E_{B}) = \sum_{i}|\braket{i|V_{PE}|g}|^{2}\delta(E_{B}+E_{g}-E_{i}),
\end{align}
where $\ket{g}$ and $\ket{i}$ denote the initial state and the intermediate state, and $E_{g}$ and $E_{i}$ are their energies, respectively.
In Eq. (\ref{eq.xps}), $V_{PE}$ describes the electronic dipole emission of the photoelectrons from the core levels.
The normal XES(NXES) and XEPECS are formulated as the second-order optical process and are calculated using the spectral function as
\begin{align}
\label{eq.nxes}
    F_{\mathrm{NXES}}(\omega) = \int dE_{B}\sum_{f}\bigg|\sum_{i}&\frac{\braket{f|V_{ph}|i}\braket{i|V_{PE}|g}}{E_{B}+E_{g}-E_{i}+i\Gamma}\bigg|^{2}\notag\\
    &\times\delta(E_{B}+E_{g}-\omega-E_{f}),\\
\label{eq.xepecs}
    F_{\mathrm{XEPECS}}(E_{B},\omega) = \sum_{f}\bigg|\sum_{i}&\frac{\braket{f|V_{ph}|i}\braket{i|V_{PE}|g}}{E_{B}+E_{g}-E_{i}+i\Gamma}\bigg|^{2}\notag\\
    &\times\delta(E_{B}+E_{g}-\omega-E_{f}),
\end{align}
where $\ket{f}$ denote the final state with the energy $E_{f}$, and $\Gamma$ is the inverse lifetime of the core hole. 
In addition, $V_{ph}$ represents the electronic dipole emission.
In the calculation, we consider the geometry used in our previous study\cite{RTANAKA2}, i.e., the unit wave vectors of the incident photon, photoelectron, and emitted photon are set to $\hat{k}_{\mathrm{in}} = (1, 0, 0)$, $\hat{k}_{\mathrm{PE}} = (0, 1, 0)$, and $\hat{k}_{\mathrm{out}} = (1/\sqrt{2}, 1/\sqrt{2}, 0)$, respectively.
It should be noted that the NXES spectrum in Eq. (\ref{eq.nxes}) is calculated by summing over both the photoelectron spin and the emitted photon polarization components.
On the other hand, the spectral functions in Eqs. (\ref{eq.xps}) and (\ref{eq.xepecs}) are, if necessary, resolved into the photoelectron spin $\sigma$ and emitted photon polarization $\lambda$, but these indices are omitted here for simplicity.

\section{Results and discussion}
\subsection{3$d\rightarrow\ $3$p$ and 3$d\rightarrow\ $2$p$ SPR-XEPECS process for $Ti_{2}O_{3}$-type system}
\begin{figure}[H]
    \centering
    \includegraphics[scale=0.6]{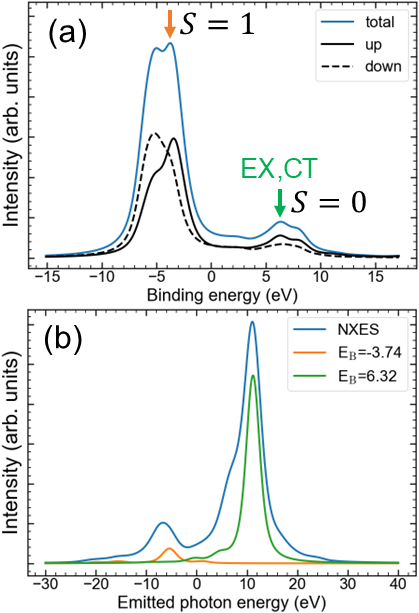}
    \caption{(a) The spin-resolved 3$p$ XPS, (b) the 3$d\rightarrow\ $3$p$ NXES and XEPECS spectra of $\mathrm{Ti_{2}O_{3}}$-type system.
    The blue curve in (b) represents the binding energy-integrated NXES spectrum, while the other spectra show the 3$d\rightarrow\ $3$p$ XEPECS at specific binding energies $E_{B} = -3.74$ eV (orange) and $6.32$ eV (green).
    Each binding energy corresponds to the peak position indicated by the arrow of the same color in (a).}
    \label{Fig.2}
\end{figure}

To investigate the dependence of spin and polarization entanglement on the excited core-level, we consider the cases of the 3$d\rightarrow\ $3$p$ and 3$d\rightarrow\ $2$p$ SPR-XEPECS processes for $\mathrm{Ti_{2}O_{3}}$-type system.
The results for the 3$d\rightarrow\ $2$p$ process have been reported in our previous work\cite{RTANAKA2}, and here we focus on the results for the 3$d\rightarrow\ $3$p$ process.
Fig. \ref{Fig.2}(a) shows the calculated results of the spin-resolved 3$p$ XPS, and Fig. \ref{Fig.2}(b) the 3$d\rightarrow\ $3$p$ NXES and XEPECS.
The continuous spectra in Fig. \ref{Fig.2}(a) are obtained by convoluting the line spectra from Eq. (\ref{eq.xps}) with the Lorentzian and Gaussian functions of the half-width $\Gamma_{\mathrm{L}} = 0.7$ eV and $\Gamma_{\mathrm{G}} = 0.5$ eV at half maxima (HWHM), respectively.
On the other hand, the spectra in Fig. \ref{Fig.2}(b) are obtained by convoluting the line spectra from Eqs. (\ref{eq.nxes}) and (\ref{eq.xepecs}) with $\Gamma_{\mathrm{L}} = 1.0$ eV and $\Gamma_{\mathrm{G}} = 1.0$ eV. 
Here, the inverse lifetime of the 3$p$ core-hole $\Gamma_{3p}$ is set to be $0.5$ eV.
Note that in the calculation of Fig. \ref{Fig.2}(b), we take the sum over the spin and polarization both for XEPECS and NXES.

In Fig. \ref{Fig.2}(a), the 3$p$ XPS shows the two-peak structure with the broad main peak structure on the low-energy side due to the multiplet effect, and a satellite formed by the overlap of the satellites labeled “EX” and “CT”.
As discussed in ref. [20], the satellite labeled “CT” corresponds to the charge-transfer (CT) satellite.
In contrast, the satellite labeled “EX” is called the exchange (EX) satellite, arising from the exchange interaction between the 3$d$ and 3$p$ electrons, which is described by the Slater integrals $G^{k}(3p,3d)$.
The main peak is formed when the spin of the 3$d$ electrons is parallel to that of the 3$p$ core electrons, corresponding to the spin-triplet state with $S=1$, whereas the EX satellite is formed when their spins are antiparallel corresponding to the spin-singlet state with $S=0$.
In Fig. \ref{Fig.2}(b), the 3$d\rightarrow\ $3$p$ NXES shows the two-peak structure with the main peak on the high-energy side.
In the calculation of the 3$d \rightarrow\ $ 3$p$ NXES shown in Fig. 2(b), we used the binding energy $E_B$ from $-15$ eV to $20$ eV at intervals of 0.2 eV. 
For each $E_B$ value, the 3$d \rightarrow\ $ 3$p$ XEPECS intensity was computed using Eq. (\ref{eq.xepecs}), and then the NXES spectrum was obtained by integrating these results according to Eq. (\ref{eq.nxes}).
In Fig. \ref{Fig.2}(b), the orange and green lines represent the XEPECS spectra calculated at the binding energies indicated by the arrows of the same colors in the 3$p$ XPS in Fig. \ref{Fig.2}(a).
Here, we note that the XEPECS intensity becomes the highest at $E_B$ = 6.32 eV among the XEPECS spectra shown in Fig. \ref{Fig.2}(b). 
This can be simply interpreted as the difference in the 3$d-$3$p$ electronic configurations between the main peak and satellite in the 3$p$ XPS of Fig. \ref{Fig.2}(a).
In the main peak, a 3$d$ up-spin electron is accompanied by a 3$p$ down-spin hole, whereas in the satellite by a 3$p$ up-spin hole.
Then, the former is dipole-forbidden, while the latter is dipole-allowed, leading to the highest XEPECS intensity at the satellite of the 3$p$ XPS.

\begin{figure}[H]
    \centering
    \includegraphics[scale=0.39]{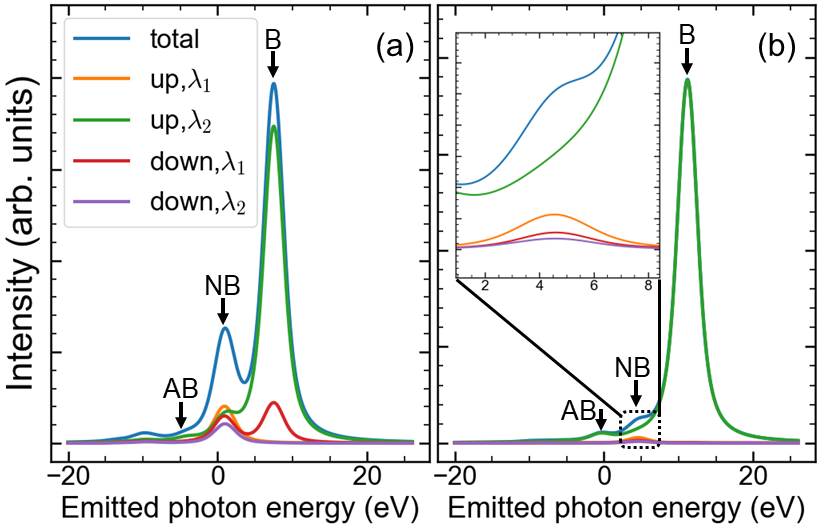}
    \caption{(a) 3$d\rightarrow\ $2$p$ SPR-XEPECS at $E_{B}$ = 2.64 eV, for 2$p$ XPS in Fig. 2(a) of ref. [17].
    (b) 3$d\rightarrow\ $3$p$ SPR-XEPECS at $E_{B}$ = 6.32 eV in the 3$p$ XPS of Fig. \ref{Fig.2}(a)}
    \label{Fig.3}
\end{figure}

Fig. \ref{Fig.3}(b) shows the  3$d\rightarrow\ $3$p$ SPR-XEPECS calculated at $E_B$ = 6.32 eV in Fig. \ref{Fig.2}(b), and Fig. \ref{Fig.3}(a) shows the 3$d \rightarrow 2$p SPR-XEPECS at $E_B$ = 2.64 eV, determined from the 2$p$ XPS result in the previous paper\cite{RTANAKA2}.
There are four decomposition patterns in the SPR-XEPECS spectrum corresponding to the combinations of the photoelectron spin (up/down) and the linear polarization of the emitted photon ($\lambda_{1}$/$\lambda_{2}$).
Here, the $z$ axis is taken as the quantization axis for the photoelectron spin (up/down), and $\lambda_{1}$ and $\lambda_{2}$ represent the horizontal and vertical linear polarizations with respect to the $xy$ plane, respectively\cite{RTANAKA2}.

In the XEPECS spectra shown in Fig. \ref{Fig.3}, the peak “B”, satellite “NB”, and “AB” correspond to the bonding, non-bonding, and anti-bonding states in the XEPECS final state, which are formed due to the hybridization between Ti 3$d$ and O 2$p$ orbitals\cite{RTANAKA2,MATSUBARA2}.
As discussed in ref. [17], in Fig. \ref{Fig.3}(a) for the 2$p$ core excitation, the main peak labeled “B” has the peak intensity corresponding to the up/$\lambda_{2}$ and down/$\lambda_{1}$, suggesting the occurrence of spin and polarization entanglement.
In fact, by evaluating the degree of entanglement using the tangle defined in Eq. (\ref{eq.tangle}), we obtained a value of 0.14, by which we have already confirmed that the spin and polarization entanglement occurs at peak “B”.
Here, these combination between the photoelectron's spin and the emitted X-ray photon's polarization can be obtained from the dipole transition amplitude calculated in our previous study\cite{RTANAKA2}.
In contrast, in Fig. \ref{Fig.3}(b) for the 3$p$ excitation, the main peak at $\omega = 11.18$ eV, labeled “B”, shows the contribution only from the up/$\lambda_{2}$, suggesting the formation of a pure state.
This arises because, as will be discussed in detail later, among the two satellites labeled “EX” and “CT” in Fig. \ref{Fig.2}(a), the “EX” appears prominently when the photoelectron has up spin, while the down-spin satellite is suppressed in the EX mechanism.
Actually, in the 3$p$ XPS of the $\mathrm{Ti^{3+}}$ ionic model, only the up-spin intensity appears in the satellite, and thus, the down-spin intensity indicated by the green arrow in Fig. \ref{Fig.2}(a) originates from the CT mechanism.
The CT down spin component further decays into the satellite “NB” of the 3$d\rightarrow\ $3$p$ XEPECS, as shown in the inset of Fig. \ref{Fig.3}(b).

\begin{figure}[H]
    \centering
    \includegraphics[scale=0.46]{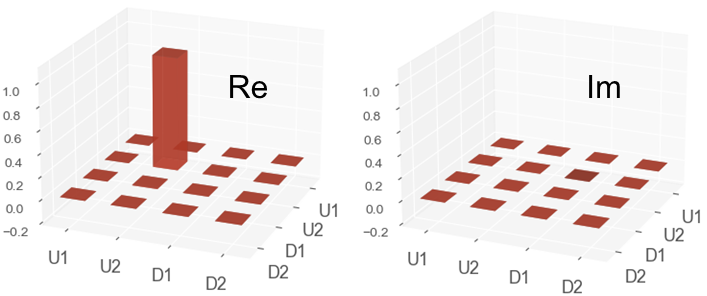}
    \caption{Real (Re) and imaginary (Im) parts of the density matrix calculated at the peak “B” in Fig. \ref{Fig.3}(b). The basis states are defined as products of the photoelectron spin (U: up, D: down) and the emitted photon's linear polarization (1: $\lambda_1$, 2: $\lambda_2$).}
    \label{Fig.4}
\end{figure}

To verify whether the quantum state is a pure state, we calculate the density matrix given in Eq. (\ref{eq.rho}) at $E_B$ = 6.32 eV and $\omega$ = 11.18 eV, i.e., at the peak “B” in Fig. \ref{Fig.3}(b). 
Fig. \ref{Fig.4} shows the real and imaginary parts of the density matrix.
The density matrix is constructed in the basis of the coupled spin–polarization states $\ket{\sigma\lambda}$, where U and D denote the photoelectron spin $\sigma$ up and down, and 1 and 2 the linear polarization $\lambda_{1}$ and $\lambda_{2}$ of the emitted X-ray photon.
As shown in Fig. \ref{Fig.4}, the density matrix clearly exhibits the diagonal element $\ket{U2}\bra{U2}$. 
As can be seen at the peak “B” of the 3$d \rightarrow$ 3$p$ SPR-XEPECS in Fig. \ref{Fig.3}(b), the density matrix also confirms the presence of a pure state corresponding to the up/$\lambda_{2}$ combination.
In fact, the fidelity $F$ with the ideal state $\ket{\psi} = \ket{U2}$ takes the value 0.99, indicating that an almost pure state is realized.
In addition, the calculated value of the tangle, the degree of entanglement, is close to 0.
Thus, these results show that spin and polarization entanglement does not occur in the 3$p$ core-level excitation process,
in contrast to the case of the 2$p$ core-level excitation.
This fact means that the generation of entanglement strongly depends on the excited core-level.

Now, in the 3$p$ core-level excitation process, we discuss why the spin and polarization entanglement does not occur, i.e., the formation of the pure state corresponding to the (up/$\lambda_{2}$) combination, using the 3$p$XPS final state of the $\mathrm{Ti^{3+}}$ ionic model for simplicity.
Here, we focus on the 3$p$ XPS final state with $S = 0$, where the EX satellite is observed, and proceed with the discussion.
First, as mentioned in our previous study\cite{RTANAKA2}, in the ground state of the $\mathrm{Ti^{3+}}$ ion, the 3$d$ electron spin is almost aligned to the up-spin because of the crystal-field effect, and the state can be represented as $\ket{g}_{\mathrm{ion}} = \ket{3d_{-1\uparrow}}$ with orbital magnetic quantum number $m=-1$ and spin $\sigma=\uparrow$.
Thus, in the 3$p$ XPS final state with $S = 0$, the photoelectron spin is fixed to up because of the dipole selection rule with $\Delta S = 0$.
Hence, the transition process from the ground state $\ket{g}_{\mathrm{ion}}$ to the 3$p$ XPS final state can be approximately expressed as follows:
\begin{align}
\label{eq:3pXPS}
    \ket{g}_{\mathrm{ion}} &\rightarrow \ket{\underline{p}d_{(S=0,M=0)}} \ket{\uparrow}_{\mathrm{el}}\\
\label{eq:spin-singlet_3d}
    &\simeq \frac{1}{\sqrt{2}}\left( \ket{\underline{3p_{\uparrow}}}\ket{3d_{\uparrow}} - \ket{\underline{3p_{\downarrow}}}\ket{3d_{\downarrow}} \right)\ket{\uparrow}_{\mathrm{el}}.
\end{align}
Here, $\ket{\underline{p}d_{(S=0,M=0)}}$ represents the spin-singlet state with $S=0$, composed of the 3$p$ core-hole $\underline{p}$ and the 3$d$ electron, and is approximately expressed in Eq. (\ref{eq:spin-singlet_3d}).
Note that in the present geometry of the XEPECS process, when the photoelectron spin is up, the total magnetic quantum number becomes $M = 0$, as shown in our previous study\cite{RTANAKA2}.
In the case of $M=0$, when the 3$p$ XPS final state shown in Eq. (\ref{eq:3pXPS}) undergoes the 3$d\rightarrow\ $3$p$ radiative decay, a photon with linear polarization $\lambda_{2}$ is emitted, as shown at the peak ``B'' in Fig. \ref{Fig.3}(b).
As a result, the pure up/$\lambda_{2}$ state is formed.
Finally, the mechanism for the formation of the pure (up/$\lambda_{2}$) state can be interpreted as follows.
The spin of the 3$d$ electron, which is aligned to the up-spin direction by the crystal-field effect, is transferred to the photoelectron spin through the strong 3$d-$3$p$ exchange interaction, while the orbital $m=-1$ of the 3$d$ electron is transferred to the linear polarization of the emitted photon in the present geometry.
As a result, the pure (up/$\lambda_{2}$) state is formed.
Here, we emphasize that the disappearance of the entanglement is not due to the formation of the mixed state but due to the formation of the pure state.

In contrast to the 3$p$ core-level excitation process, in the 2$p$ excitation process, the spin--polarization entangled state composed of (up/$\lambda_{2}$) and (down/$\lambda_{1}$) is formed, as shown in our previous study\cite{RTANAKA2}.
This formation can be qualitatively explained by using the 2$p$ XPS final state in the $\mathrm{Ti^{3+}}$ ionic model.
In the 2$p$ excitation process, the 3$d-$core exchange interaction is weaker than that in the 3$p$ excitation process, and under such conditions, the 2$p$ XPS final state can be approximately expressed as follows:
\begin{align}
\label{eq:2p_XPS}
    \ket{3d_{-1\uparrow}}\otimes\left( \ket{\underline{2p_{\uparrow}}}\ket{\uparrow}_{\mathrm{el}} + \ket{\underline{2p_{\downarrow}}}\ket{\downarrow}_{\mathrm{el}} \right).
\end{align}
The first term in Eq. (\ref{eq:2p_XPS}), $\ket{3d_{-1\uparrow}}\ket{\underline{2p_{\uparrow}}}\ket{\uparrow}_{\mathrm{el}}$, is allowed as a radiative decay because the spins of the 3$d$ electron and the 2$p$ core-hole are parallel, while $\ket{3d_{-1\uparrow}}\ket{\underline{2p_{\downarrow}}}\ket{\downarrow}_{\mathrm{el}}$ is forbidden.
However, the 2$p$ excitation process, the spin of the 2$p$ core-hole can flip by the strong 2$p$ spin-orbit interaction. 
Therefore, the second term also contributes to the radiative decay.
Actually, the calculated 2$p$ core-hole spin occupancy is ($\braket{\underline{2p_{\uparrow}}}$, $\braket{\underline{2p_{\downarrow}}}$) = (0.35, 0.65) when the photoelectron spin is down\cite{RTANAKA2}.
Thus, the spin--polarization entangled state is formed by the spin flip of the 2$p$ core-hole due to the 2$p$ spin-orbit interaction.

\subsection{4$f\rightarrow\ $4$d$ SPR-XEPECS process for $CeF_{3}$-type system}
\begin{figure}[H]
    \centering
    \includegraphics[scale=0.65]{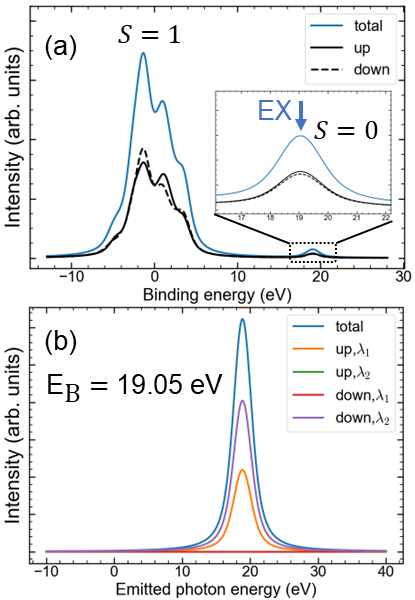}
    \caption{(a) The spin-resolved 4$d$ XPS of $\rm CeF_3$. 
    (b) The 4$f \rightarrow\ $ 4$d$ SPR-XEPECS of $\rm CeF_3$ at the binding energy $E_B = 19.05$ eV indicated by the blue arrow in the inset of (a).}
    \label{Fig.5}
\end{figure}

Figs. \ref{Fig.5}(a), \ref{Fig.5}(b) show the calculated results of spin-resolved 4$d$ XPS, 4$f\rightarrow\ $4$d$ SPR-XEPECS for $\rm CeF_{3}$-type system.
In the continuous spectrum shown in Fig. \ref{Fig.5}(a), the Lorentzian and Gaussian functions of the half-width $\Gamma_{\rm L} = 0.7$ eV and $\Gamma_{\rm G} = 0.5$ eV is assumed in Eq. (\ref{eq.xps}).
In Fig. \ref{Fig.5}(b), the continuous spectrum is obtained, as in the 3$d\rightarrow\ $3$p$ XEPECS calculation for $\rm Ti_2O_3$, by convoluting the line spectra calculated from Eq. (\ref{eq.xepecs}) with $\Gamma_{\rm L} = \Gamma_{\rm G} = 1.0$ eV. 
In addition, the inverse lifetime of the 4$d$ core-hole $\Gamma_{4d}$ is set to be 0.5 eV.

In the 4$d$ XPS spectrum shown in Fig. \ref{Fig.5}(a), as in the case of the 3$p$ XPS of $\mathrm{Ti_{2}O_{3}}$, the main peak corresponding to the spin-triplet state ($S=1$) appears on the low-binding-energy, while the EX satellite corresponding to the spin-singlet state ($S=0$) on the high-binding-energy.
In contrast to the case of $\rm Ti_{2}O_{3}$, however, the EX satellite appears in both up- and down-spin intensities, as shown in the inset of Fig. \ref{Fig.5}(a).
This fact can be explained by using a 4$d$ XPS final state with $S=0$ in a similar manner to the 3$p$ core excitation process in $\mathrm{Ti_{2}O_{3}}$.
First, we describe the ground state in $\mathrm{CeF_{3}}$.
As mentioned above, $\mathrm{CeF_{3}}$ is treated as a free $\mathrm{Ce^{3+}}$ ionic model without the crystal-field effect, and thus
the ground state $\ket{g}_{\rm ion}$ in $M=-1/2$ subspace becomes a superposition of up- and down-spin components due to the 4$f$ spin-orbit interaction:
\begin{align}
\label{eq:g_ce}
    \ket{g}_{\rm ion} = \frac{1}{\sqrt{7}} ( \sqrt{3} \ket{4f_{0\downarrow}} - 2 \ket{4f_{-1\uparrow}} ).
\end{align}
Thus, unlike the 3$p$ XPS final state with $S = 0$ shown in Eq. (\ref{eq:3pXPS}), the 4$d$ XPS final state has both up- and down-spin components of the photoelectron because the 4$f$ electron spin is transferred to the photoelectron spin according to the dipole selection rule ($\Delta S = 0$).
The transition process in the present geometry from such a ground state to the 4$d$ XPS final state can be expressed in a restricted form as follows:
\begin{align}
\label{eq:4dXPS}
    \ket{g}_{\mathrm{ion}} &\rightarrow\sum_{m=(-1,1)}A_{m}\ket{\underline{d}f_{(S=0,M=m)}} \ket{\uparrow}_{\mathrm{el}} + A_{0}\ket{\underline{d}f_{(S=0,M=0)}}\ket{\downarrow}_{\mathrm{el}}.
\end{align}
Here, $\ket{\underline{d}f_{(S=0,M=\pm1)}}$ represents the spin-singlet states with the total magnetic quantum number $M = \pm1$, and $\ket{\underline{d}f_{(S=0,M=0)}}$ represents the spin-singlet state with $M = 0$.
In addition, the coefficients $A_{\pm1}$ and $A_{0}$ represent the weights of the corresponding states, which depends on the geometry, as in the previous paper\cite{RTANAKA2}.
These $M$ values can be confirmed by calculating the dipole transition matrix $\braket{i|V_{PE}|g}$ for the 4$d$ core-level excitation process, as in our previous study\cite{RTANAKA2}, where $\ket{i}$ represents the 4$d$ XPS final state.
In this case with such $M$ values, by calculating the dipole transition matrix between the final state corresponding to the 4$f\rightarrow\ $4$d$ XEPECS process and the $\ket{\underline{d}f_{(S=0,M=m)}}$ basis states, it is found that the matrix elements corresponding to only the pair of the up-spin and polarization $\lambda_{1}$ and the pair of the down-spin and $\lambda_{2}$ have finite values.

\begin{figure}[H]
    \centering
    \includegraphics[scale=0.45]{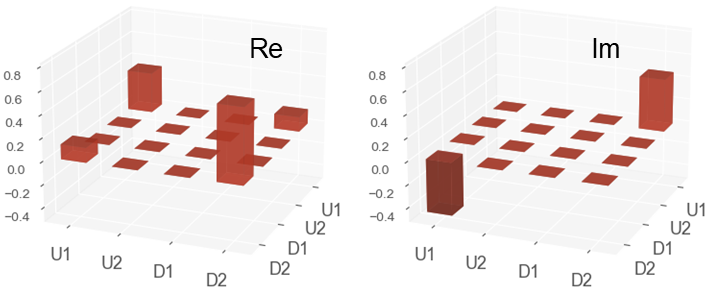}
    \caption{Real (Re) and imaginary (Im) parts of the density matrix calculated at the main peak in Fig. \ref{Fig.5}(b). The basis states are defined as products of the photoelectron spin (U: up, D: down) and the emitted photon’s linear polarization (1: $\lambda_{1}$, 2: $\lambda_{2}$).}
    \label{Fig.6}
\end{figure}

Fig. \ref{Fig.5}(b) shows the 4$f\rightarrow\ $4$d$ SPR-XEPECS at the energy corresponding to the EX satellite in Fig. \ref{Fig.5}(a), i.e., at the binding energy $E_B = 19.05$ eV. 
The main peak located at $\omega = 18.88$ eV in Fig. \ref{Fig.5}(b) is observed only for the spin and polarization pairs of up/$\lambda_{1}$ and down/$\lambda_{2}$, as discussed above.
To evaluate the spin and polarization entanglement, we show the density matrix in Fig. \ref{Fig.6} calculated at $E_B = 19.05$ eV and $\omega = 18.88$ eV.
As shown in Fig. \ref{Fig.6}, the presence of the off-diagonal elements $\ket{U1}\bra{D2}$ and $\ket{D2}\bra{U1}$ indicates the coherence between $\ket{U1}$ and $\ket{D2}$, i.e., the generation of spin and polarization entanglement.
Assuming the maximally entangled state
\begin{align}
    \ket{\psi} = \frac{1}{\sqrt{2}} ( \ket{U1} + e^{i\alpha} \ket{D2} ),
\end{align}
the fidelity is evaluated as $F = 0.98$ at the phase factor $\alpha = 4.98$.
This high fidelity value well exceeds the classical threshold of 0.5, confirming the generation of strong spin and polarization entanglement.
In addition, the tangle, which quantifies the degree of entanglement, reaches $T = 0.91$ in the present calculation.
These results show that a nearly maximally entangled state is formed at the main peak. 
This strong entanglement originates from combined conditions: (\romani) absence of the crystal field, (\romanii) existence of $S=0$ satellite in the $4f^{1}$ system, and (\romaniii) specific geometry employed in the present study.

As discussed so far, there are two distinct mechanisms for the formation of quantum entanglement in X-ray inner-shell excitation.
The first one is the 3$d\rightarrow\ $2$p$ SPR-XEPECS process of $\mathrm{Ti_{2}O_{3}}$, where the spin and polarization entanglement is generated by the spin-flip effect associated with the spin-orbit interaction of the 2$p$ core electrons.
The second one is the 4$f \rightarrow$ 4$d$ process of $\mathrm{CeF_{3}}$,
where the entangled state is formed because the dipole selection rule at the $S=0$ satellite transfers the spin component of the $4f^{1}$ spin--orbital entangled state to the spin of the photoelectron, while the present geometry transfers its orbital component to the polarization of the emitted X-ray photon.
On the other hand, the transfer mechanism of the valence electron information to the photoelectron at $S=0$ excitation results in the pure state in the $3d^{1}$ system.
In summary, the former arises from the spin-orbit interaction of the inner-shell 2$p$ electrons, whereas the latter originates from the combined effect of the spin-orbit interaction of the outer-shell 4$f$ electrons and the 4$f-$4$d$ exchange interaction.

In the present calculation, we focused on the subspace labeled by $M=-1/2$ to investigate the dependence on the excited core-level.
In actual materials, however, which of the $M=-3/2$, $1/2$, and $3/2$ case is realized---and how the corresponding states are aligned (e.g., ferromagnetically or antiferromagnetically)---depends on the material. 
These aspects are material-dependent and must be considered on a case-by-case.
As these issues go beyond the main focus of the present study, we do not examine them in detail here.
These issues are important future challenges, and we believe that a fully angle-, spin-, and polarization-resolved XEPECS calculation environment will be a powerful tool for addressing them.

\section{Conclusion}
In this study, we theoretically investigated the dependence of spin and polarization entanglement in the SPR-XEPECS process on the excited core-level. 
Specifically, we treated the 3$d\rightarrow\ $2$p$ and 3$d\rightarrow\ $3$p$ SPR-XEPECS processes in $\rm Ti_{2}O_{3}$-type system, and the 4$f\rightarrow\ $4$d$ process in $\rm CeF_{3}$-type system. 
For the $\rm Ti_{2}O_{3}$, we used a $\rm TiO_{6}$-type cluster model with the full-multiplet structure of Ti ion and with the charge-transfer effects between the Ti 3$d$ and O 2$p$ orbitals, whereas for the $\rm CeF_{3}$, we used a free $\mathrm{Ce^{3+}}$ ion with the full-multiplet structure of the Ce ion.
The degree of entanglement was evaluated using the fidelity and tangle based on the density matrix.

We found that, in the 3$d\rightarrow\ $2$p$ process of $\mathrm{Ti_{2}O_{3}}$, the spin and polarization entanglement occurs due to the strong spin-orbit interaction of the 2$p$ core electron.
In contrast, in the 3$d\rightarrow\ $3$p$ process, the pure state corresponding to the (up/$\lambda_{2}$) combination is formed, i.e., no entanglement is generated.
This is because the spin component of the 3$d$ electron, aligned in the up-spin direction by the crystal-field effect, is transferred to the photoelectron spin through the strong 3$d-$3$p$ exchange interaction, and subsequently, the orbital component of the 3$d$ electron is transferred to the emitted photon polarization $\lambda_{2}$.
In the 4$f\rightarrow\ $4$d$ SPR-XEPECS process of $\mathrm{CeF_{3}}$, strong spin--polarization entangled state composed of (up/$\lambda_{1}$) and (down/$\lambda_{2}$) is formed. 
This is because the spin and orbital entanglement of the 4$f$ electron is transferred to the spin of the photoelectron and to the polarization of the emitted photon, under the combined conditions: (\romani) absence of the crystal field, (\romanii) existence of $S=0$ satellite in the $4f^{1}$ system, and (\romaniii) specific geometry employed in the present study.

This work reveals a dependence of spin and polarization entanglement on the excited core-level, and further reveals the existence of two distinct mechanisms for entanglement generation.
One originates from the spin-orbit interaction of core electrons such as the 2$p$ electron, whereas the other originates from the spin-orbit interaction of the 4$f$ electron and the strong exchange interaction between the $4f$ and $4d$ electrons.
These findings on the excited core-level dependence clarify the characteristic features of quantum entanglement in X-ray inner-shell excitation.

\begin{acknowledgment}

We would like to thank Professor Satoshi Tanaka for valuable discussions. This work was supported by JST SPRING, Grant Number JPMJSP2139.

\end{acknowledgment}

\end{document}